\documentclass[preprint]{aastex}

\usepackage{psfig}

\newcommand{\kms}{\, {\rm km\, s}^{-1}}

\newcommand{\ergs}{\, {\rm ergs} \, {\rm s}^{-1}}
\newcommand{\cm}{\, {\rm cm}}
\newcommand{\seg}{\, {\rm s}}
\newcommand{\hz}{\, {\rm Hz}}
\newcommand{\ster}{\, {\rm sr}}

\newcommand{\kpc}{\, {\rm kpc}}

\newcommand{\lya}{Ly$\alpha$ }

\newcommand{\nhi}{N_{\rm HI}}

\begin{document}

\slugcomment{Accepted for publication in ApJ}

\shortauthors{Zheng \& Miralda-Escud\'e}
\shorttitle{Monte Carlo Simulation of \lya Scattering}

\title{Monte Carlo Simulation of \lya Scattering and Application
to Damped \lya Systems}
\author{Zheng Zheng \& Jordi Miralda-Escud\'e}
\affil{Department of Astronomy, The Ohio State University, 140 West 18th 
Avenue, Columbus, OH 43210, USA}
\email{zhengz@astronomy.ohio-state.edu, jordi@astronomy.ohio-state.edu}

%%%%%%%%%%%%%%%%%%%%%%%%%%%%%%%
\begin{abstract}

  A Monte Carlo code to solve the transfer of \lya photons is
developed, which can predict the \lya image and two-dimensional \lya
spectra of a hydrogen cloud with any given geometry, \lya emissivity,
neutral hydrogen density distribution, and bulk velocity field. We apply
the code to several simple cases of a uniform cloud to show how the \lya
image and emitted line spectrum are affected by the column density, 
internal velocity gradients, and emissivity distribution. We then apply
the code to two models for damped \lya absorption systems: a spherical,
static, isothermal cloud, and a flattened, axially symmetric, rotating
cloud. If the emission is due to fluorescence of the external background
radiation, the \lya image should have a core corresponding to the region
where hydrogen is self-shielded. The emission line profile has the
characteristic double peak with a deep central trough. We show how
rotation of the cloud causes the two peaks to shift in wavelength
as the slit is perpendicular to the rotation axis, and how the 
relative amplitude of the two peaks is changed. In reality, damped \lya 
systems are likely to have a clumpy gas distribution with turbulent 
velocity fields, which should smooth the line emission profile, but 
should still leave the rotation signature of the wavelength shift 
across the system.
 
\end{abstract}

\section{Introduction}

  The high column density absorption systems (damped \lya and Lyman
limit systems) are a key part of understanding galaxy formation. As a
galaxy collapses from the highly ionized intergalactic medium, the gas
will inevitably go through a phase of a self-shielded cloud of
atomic hydrogen before it can cool and collapse further into molecular
clouds, form stars, and increase its metallicity. Many of the damped
\lya systems (DLAs) at high redshift may be gaseous halos in the process of
forming galaxies rather than fully formed spiral disks, especially
in view of the very low metallicities at high redshifts (\citealt{Lu96,
Prochaska01}). If so, the internal structure of
these objects should reveal to us the detailed processes by which
galaxies form. Several basic questions arise in relation to this: how
big are the absorption systems? Is the gas smoothly distributed, or is
it in the form of clumps and an interclump medium? If so, how big are
the clumps? What is the dynamical state of the gas? How does its mean
rotation velocity compare to its velocity dispersion?

  Some of these questions may be addressed by studying the associated
metal absorption lines (e.g., \citealt{Prochaska97,Prochaska98}) or by
radio and optical observations of galaxies found to be associated with
the absorbers (see \citealt{Briggs89,Djorgovski96}), although the 
majority of DLAs at high redshifts might not be associated with luminous 
galaxies.

  An alternative observational probe of the structure of DLAs
and Lyman limit systems (LLSs) may be found in their \lya emission produced
in hydrogen recombinations and the subsequent scattering of \lya photons
through the gas (\citealt{Hogan87,Gould96}). The
source of the ionization may be the external cosmic ionizing radiation,
shock-heating due to the gravitational collapse of the gas in a dark
matter halo, or star formation within the system. In the case of
fluorescence of external radiation, the maximum surface brightness is
achieved in any system with $\nhi \gtrsim 10^{18}\cm^{-2}$, where all
the incident external photons are absorbed. 

  Efforts have been made to image known DLAs and LSSs towards high-$z$ 
quasars.  DLAs towards PKS0528-250 \citep{Warren96} and Q0151+048A 
\citep{Fynbo99} and one LLS towards Q1205-30 \citep{Fynbo00} have been
successfully detected using tuned narrow band filters. There are also some 
spectroscopically confirmed detections of \lya emission from DLAs 
(e.g. \citealt{Hunstead90,Djorgovski98,Pettini95,Leibundgut98}). Most of 
these absorber systems with \lya emission detected are near the redshift 
of the background quasar, which suggests a \lya source due to 
photoionization by the quasar (\citealt{Fynbo99,Fynbo00}). 

%A recent study shows that the emission 
%properties of DLA galaxies are consistent with the conjecture that they
%are Lyman-break galaxies \citep{Moller02}. In addition, some of the high-$z$ 
%\lya emitters (\lya blobs) found in blank-sky fields (\citealt{Pascarelle96,
%Cowie98,Keel99,Steidel00}) may be related to DLAs. These \lya blobs
%may originate from superwinds driven by the initial starburst in galaxies
%(\citealt{Taniguchi00,Taniguchi01}) or cooling radiation from galaxy 
%formation (e.g., \citealt{Fardal01}).

  Hydrogen \lya line is a resonance line. The problem of radiative 
transfer of resonance line radiation can be approximately solved 
analytically under certain conditions (e.g., \citealt{Harrington73,
Harrington74,Neufeld90}). These analytic solutions can be found only 
for a limited number of cases such as a static, extremely opaque and 
plane-parallel medium. In some cases, numerical methods are used to 
solve the transfer equation (e.g., \citealt{Adams72,Hummer80,Urbaniak81}).
For a more general geometry, density distribution, and kinematics, the 
Monte Carlo method turns out to be very useful and thus it has been 
applied to many problems (e.g., \citealt{Auer68,Avery68,Caroff72,
Panagia73,Bonilha79,Natta86,Ahn00,Ahn01,Ahn02}). 

  This paper presents the results of a numerical Monte Carlo code we
have developed to compute the transfer of \lya photons through an
arbitrary distribution of hydrogen. The code predicts the
two-dimensional (2-D) image and line-spectrum along any given observed
direction, for a general three-dimensional distribution of gas of
given geometry, \lya emissivity, neutral hydrogen density, and bulk
velocity field. In this paper, it will be applied only to spherical
and axially symmetric clouds, although in the future it can be applied
to results of numerical simulations of galaxy formation.

  We describe the Monte Carlo code in \S 2.
In \S 3, we apply the code to several simple cases of hydrogen clouds
to demonstrate the effects of the neutral hydrogen column density, \lya 
emissivity and bulk velocity field on the spectra of escaped \lya photons.
In \S 4, we model DLAs as static and
rotating clouds and investigate their 2-D \lya emission spectra.
Finally, we have a brief summary and discussion in \S 5.
 
\section{Monte Carlo Simulation of \lya Scattering}

\subsection{Description of the Method}

  The method used to numerically compute the images and spectra of \lya
emission consists of generating random realizations of the trajectory 
of a large number of \lya photons as they are scattered within the
specified gas distribution. In a nutshell, the method proceeds through
the following steps: First, we generate the point where the photon is
created, with the emissivity distribution proportional to the recombination
rate at every point. Second, the optical depth $\tau$ that the photon 
will travel through before it is scattered is randomly generated with 
a probability density $e^{-\tau}$. The spatial location of the scattering 
at optical depth $\tau$ from the point of emission is then determined 
along a randomly chosen direction, with the knowledge of the neutral 
hydrogen density distribution and the scattering cross section (see eq.[2]
below).  We then generate the velocity of the atom that scatters the photon, 
as described below, compute the new frequency of the photon after scattering, 
and generate the new direction of the photon. The process of propagation 
and scattering is repeated until the photon escapes the modeled system. 

  As these random realizations of photon trajectories are being carried
out, a \lya image of the system along a specified direction can be
created. The image consists of a three-dimensional array of the two
projected coordinates perpendicular to the direction of the image and
the frequency of the escaped photons. The array contains the mean number of
photons emitted at every projected position and frequency. At every 
scattering of a photon, the probability that the photon is re-emitted in 
the direction of the image and escapes is separately calculated. This 
probability is added on the element of the image array corresponding to 
the projected position and frequency of the photon. In practice, for the 
majority of the scatterings the optical depth is large and the probability 
of escape is negligible, so one can avoid the computation of the optical 
depth in the direction of the image on most scatterings.

  We now describe how the optical depth of the \lya photons is computed 
and the velocity of the atom at each scattering is generated.
The scattering cross section of \lya photons as a function of the
frequency in the rest frame of the hydrogen atom is
$$
\sigma_\nu= f_{12}\, \frac{\pi e^2}{m_e c}\, \frac{\Delta\nu_L / 2\pi}
{(\nu-\nu_0)^2+(\Delta\nu_L/2)^2} ~, \eqno(1)
$$ 
where $f_{12}=0.4162$ is the \lya oscillator strength,
$\nu_0= 2.466\times10^{15}$Hz is the line center frequency,
$\Delta\nu_L= 4.03\times 10^{-8} \nu_0 = 9.936\times10^7$ Hz is the
natural line width, and the other symbols have their usual meaning.
For a hydrogen atom with a velocity component $v_z$ along the 
photon's direction in a fixed ``laboratory'' frame, the term $\nu-\nu_0$ 
in the above equation becomes
$(\nu_i-\nu_0) - (v_z / c)\nu_0$, where $\nu_i$ is the frequency of
the incident photon in the ``laboratory'' frame. For a Maxwellian
distribution of atom velocities, the resulting average cross section is
$$ 
\sigma(x_i) = f_{12}\, \frac{\sqrt{\pi}e^2}{m_ec\, \Delta\nu_D}
H(a,x_i) ~, \eqno(2)
$$
where 
$$H(a,x) = \frac{a}{\pi}\, \int_{-\infty}^{+\infty}
\frac{e^{-y^2}}{(x-y)^2+a^2}\, dy \eqno(3)
$$ 
is the Voigt function, $\Delta\nu_D= (v_p /c)\, \nu_0$ is the Doppler 
frequency width, $v_p= (2kT/m_H)^{1/2}$ is the atom velocity dispersion 
times $\sqrt{2}$, $T$ is the gas temperature, $m_H$ is the mass of the 
hydrogen atom, $x_i= (\nu_i-\nu_0)/\Delta\nu_D$ is the relative frequency 
of the incident photon in the laboratory frame, and
$a=\Delta\nu_L/(2\Delta\nu_D)$ is the relative line width. The optical
depth along the photon trajectory is computed by integrating the cross
section in equation (2) times the atomic hydrogen density.

  Once a spatial location where the photon is scattered has been 
determined, the velocity $v_z$ along the direction of the incident photon 
of the atom responsible for the scattering obeys the following 
distribution:
$$
f(u_z) = \frac{a}{\pi}\, \frac{e^{-u_z^2}}{(x_i-u_z)^2+a^2}\,
H^{-1}(a,x_i) ~,  \eqno(4)
$$
where $u_z=v_z/v_p$. We describe how we generate random numbers with this
distribution in the Appendix. The two velocity components perpendicular to 
the direction of the photon simply follow a Gaussian distribution
parameterized by the local temperature. 

  After the total velocity of the atom is chosen according to the 
above distribution, we first perform a Lorentz transform 
of the direction and frequency of the photon to the rest frame of the 
atom. The direction of the scattered photon is then generated according 
to a dipole distribution in this frame. Its frequency differs from the 
incident one by the recoil effect which is taken into account in the 
code (it is negligibly small in the applications we will present in this 
paper). The direction and frequency of the scattered photon is then 
transformed back to the laboratory frame, a new optical depth is chosen, 
and the entire process is repeated until the photon escapes from the cloud. 
In the presence of a fluid velocity, the procedure changes only by replacing 
the incident frequency $x_i$ in equations (2) and (4) by its value in the 
fluid frame, $x_{{\rm f}i}= x_i-(v_{{\rm f}z}/c)(\nu_0/\Delta\nu_D)$, 
where $v_{{\rm f}z}$ is the fluid velocity parallel to the photon's direction 
(valid in the non-relativistic regime).

  As mentioned previously, the \lya image and spectrum of the system are 
calculated as the photon trajectories are randomly generated. At every
photon scattering, we compute first the optical depth for the photon
to escape the cloud along the direction of the image, and the photon
is added to the corresponding element of the image array in projected
position and frequency, with the weight $e^{-\tau} (1+\mu^2) {\rm d} \Omega$,
where $\tau$ is the optical depth for escaping, $\mu$ is the cosine
of the angle between the incident photon and the image direction, and 
${\rm d} \Omega$ is the solid angle subtended by the pixel (both
$\mu$ and ${\rm d} \Omega$ are measured in the rest frame of the atom). 
The factor $1+\mu^2$ accounts for the dipole probability distribution of
the direction of the photon after scattering. A photon also has a probability 
to directly escape the system along the direction of the image when it is 
created which is added to the corresponding element of the image 
array. After the contribution to the image has been added, a random 
direction of the new scattered photon is generated according to the same 
dipole distribution, and the realization of the photon trajectory is 
continued until the photon finally escapes from the cloud. The total number
of photons in the simulation is chosen to give an acceptable signal-to-noise 
ratio (the typical number of photons used for simulations in this paper is 
on the order of $\sim 10^3-10^4$). 
 
\subsection{Tests of the Code}

\begin{figure}[h]
\centerline{\psfig{figure=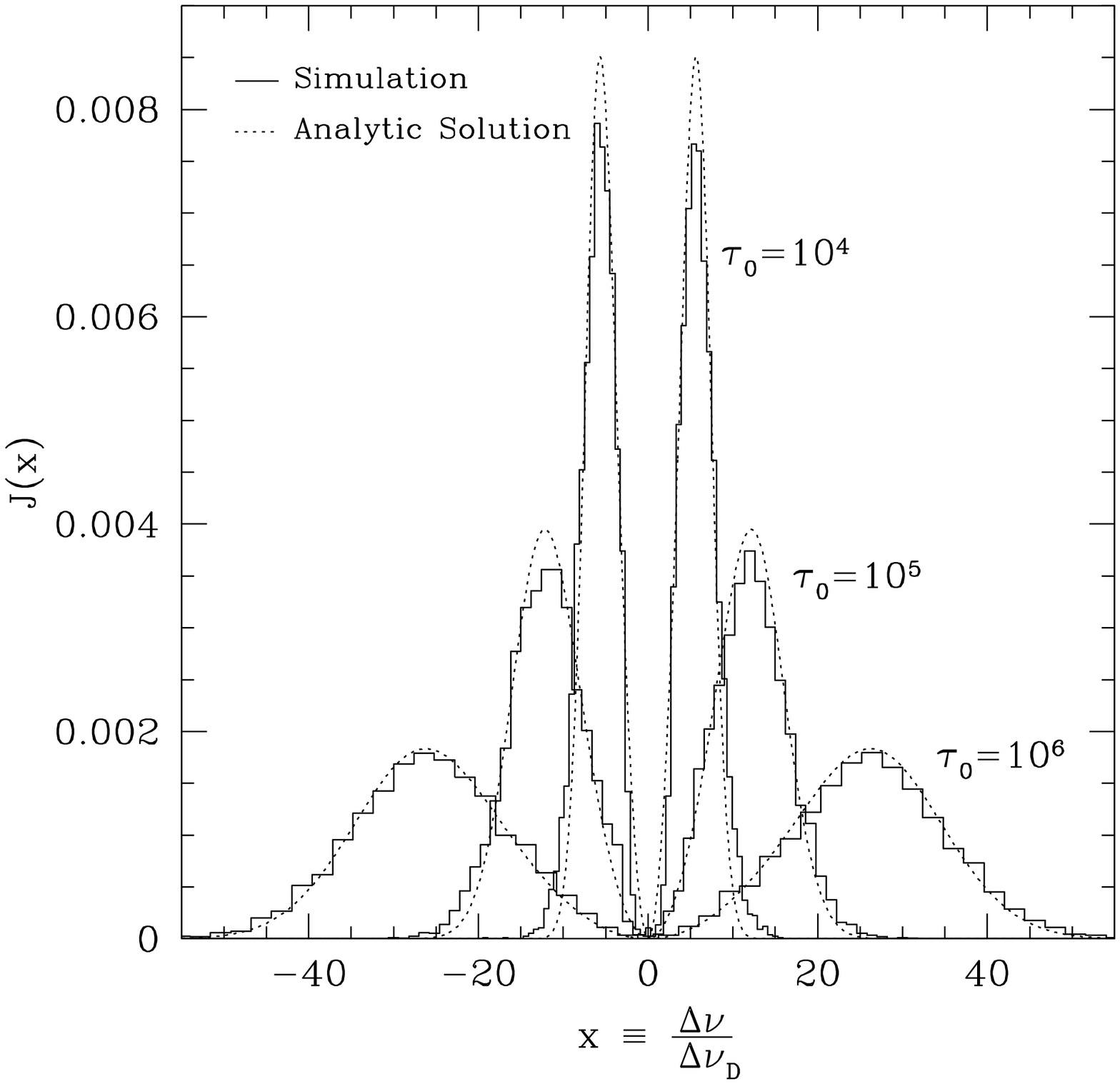,height=8cm,width=8cm}
            \psfig{figure=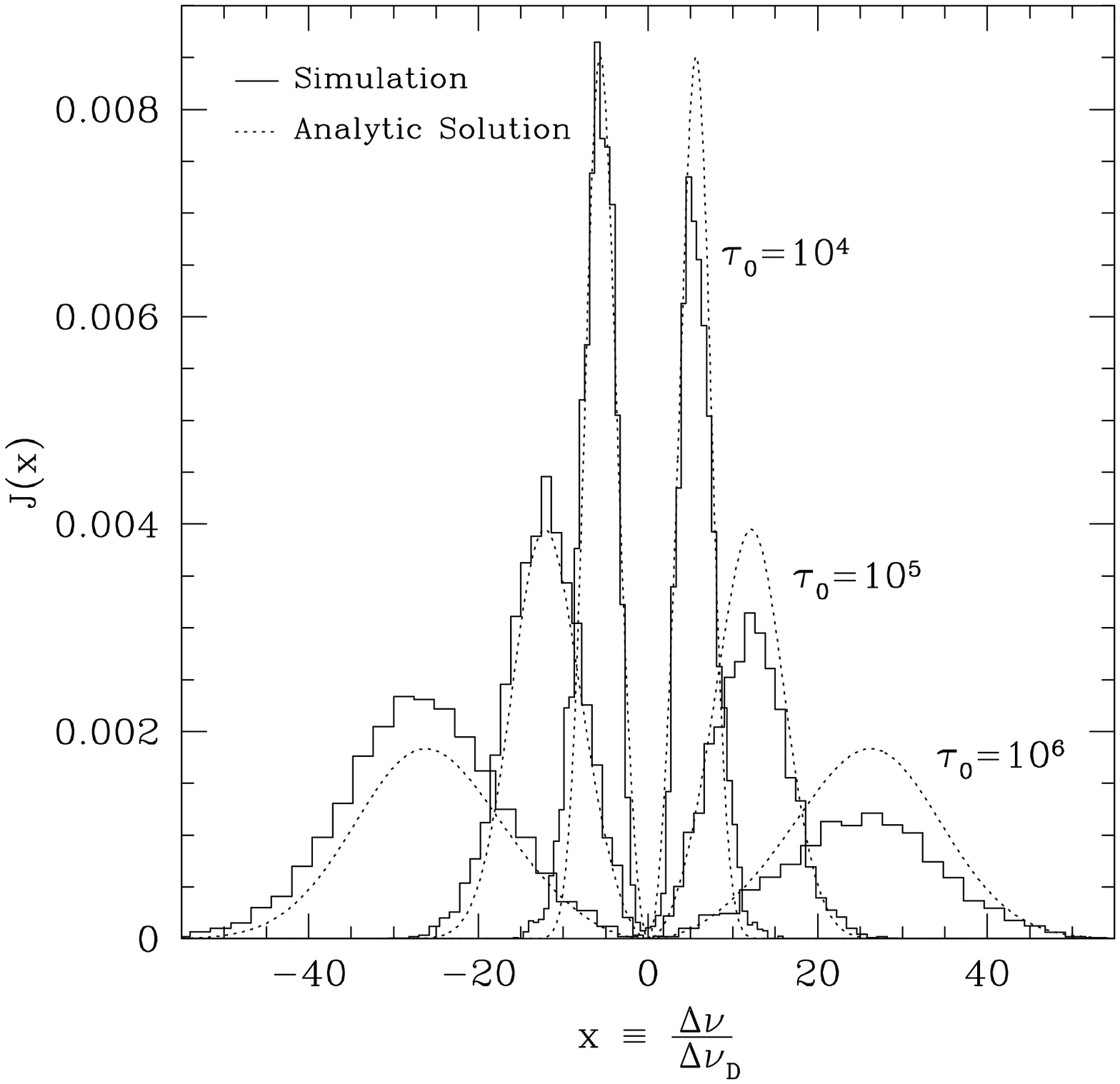,height=8cm,width=8cm}
           }
\caption[]{\label{fig:test}
Comparison between results from our Monte Carlo simulations and
the analytic solutions. Dotted lines are the analytic solutions
(Neufeld 1990) of the \lya spectra at one boundary of a slab
with a midplane source and different scattering optical depths.
Solid lines are those from simulations. In the left panel, the 
recoil effect is neglected in the simulations; in the right panel,
this effect is taken into account in the simulations (see the text). 
}
\end{figure}

  We test our numerical code for the case of a static, plane-parallel
slab, for which \citet{Neufeld90} derived an analytic solution of the 
mean intensity in the limit of a large optical depth. We simulate the 
case of a midplane source radiating 
line-center \lya photons in a plane-parallel slab of uniform density 
for three optical depths: $\tau_0=10^4, 10^5,$ and $10^6$, where 
$\tau_0=\sigma(x_i=0) \nhi$ is the line-center optical depth from the 
midplane to the boundary of the slab. We assume a temperature $T=10K$ 
($a \approx 1.49 \times 10^{-2}$). As we have mentioned, the
recoil effect is taken into account in the code. For the purpose of 
comparison, we first turn off this effect and run the simulation. 
We compare the spectra of the escaped \lya photons with the analytic 
results (eq.[2.24] in \citealt{Neufeld90}; note that his definition 
of $\tau_0$ differs from ours by a factor of $\sqrt{\pi}$ as $a \ll 1$. 
See also \citealt{Ahn01}). The left panel of Figure \ref{fig:test} 
shows excellent agreement, becoming better as the optical depth increases 
as expected since the analytic solution applies in the limit of a large 
optical depth ($\sqrt{\pi}\tau_0 \gtrsim 10^3/a$, \citealt{Neufeld90}). 
We then turn on the recoil effect and rerun the simulation.
The results are shown in the right panel of Figure \ref{fig:test}.  
The double-peaked frequency distribution of the escaped photons 
becomes asymmetric -- more photons appear at the red part.  The profile 
from analytic solutions can be regarded as an average of the red part and 
the blue part. The magnitude of the recoil effect is easily understood as 
reflecting the thermalization of photons around frequency $\nu_0$ inside 
the cloud (\citealt{Wouthuysen52, Field59}), which modifies their abundance
by a factor of $e^{-x/x_T}$, where $x\equiv \Delta\nu/\Delta\nu_D$ and  
$x_T\equiv kT/(h\Delta\nu_D)=(m_Hc^2kT/2)^{1/2}/(h\nu_0)$. 
Obviously, the recoil effect becomes more important for low temperature and
large optical depth.

  We also test the code for a spherically-symmetric case with a particular 
bulk velocity field. \citet{Loeb99} study and simulate the scattered \lya 
radiation around a point source before cosmological reionization. The \lya 
photons are scattered by the intergalactic medium (IGM) which is undergoing 
Hubble expansion. According to their approximation, the gas temperature is 
regarded as zero. We run a corresponding simulation with our code and find 
that the frequency distribution and the surface brightness profile of the 
escaped \lya photons shown in \citet{Loeb99} are well reproduced.

\section{Spherical Clouds of Uniform Density}

\begin{figure}[h]
\centerline{\psfig{figure=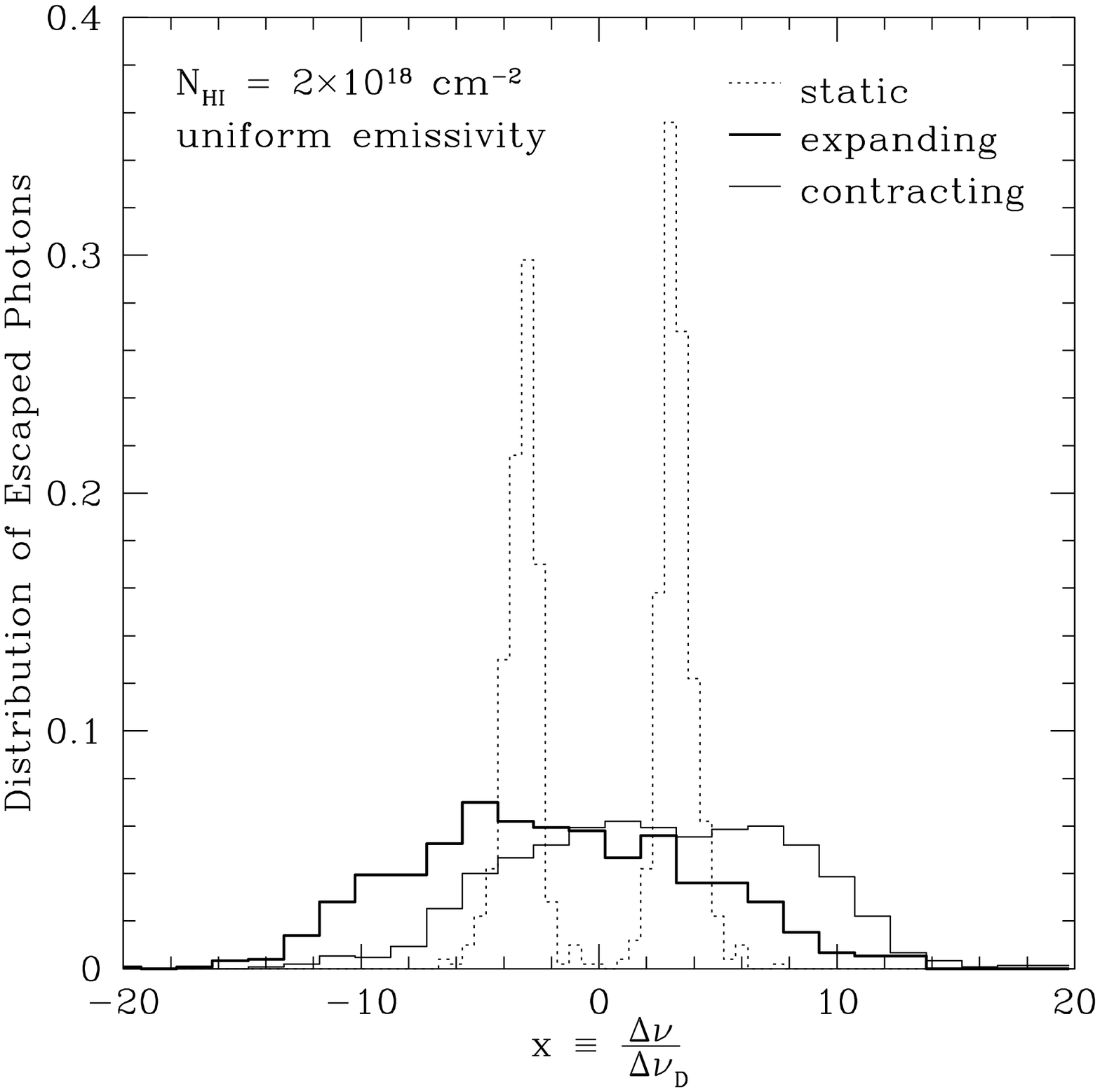,height=8cm,width=8cm}
            \psfig{figure=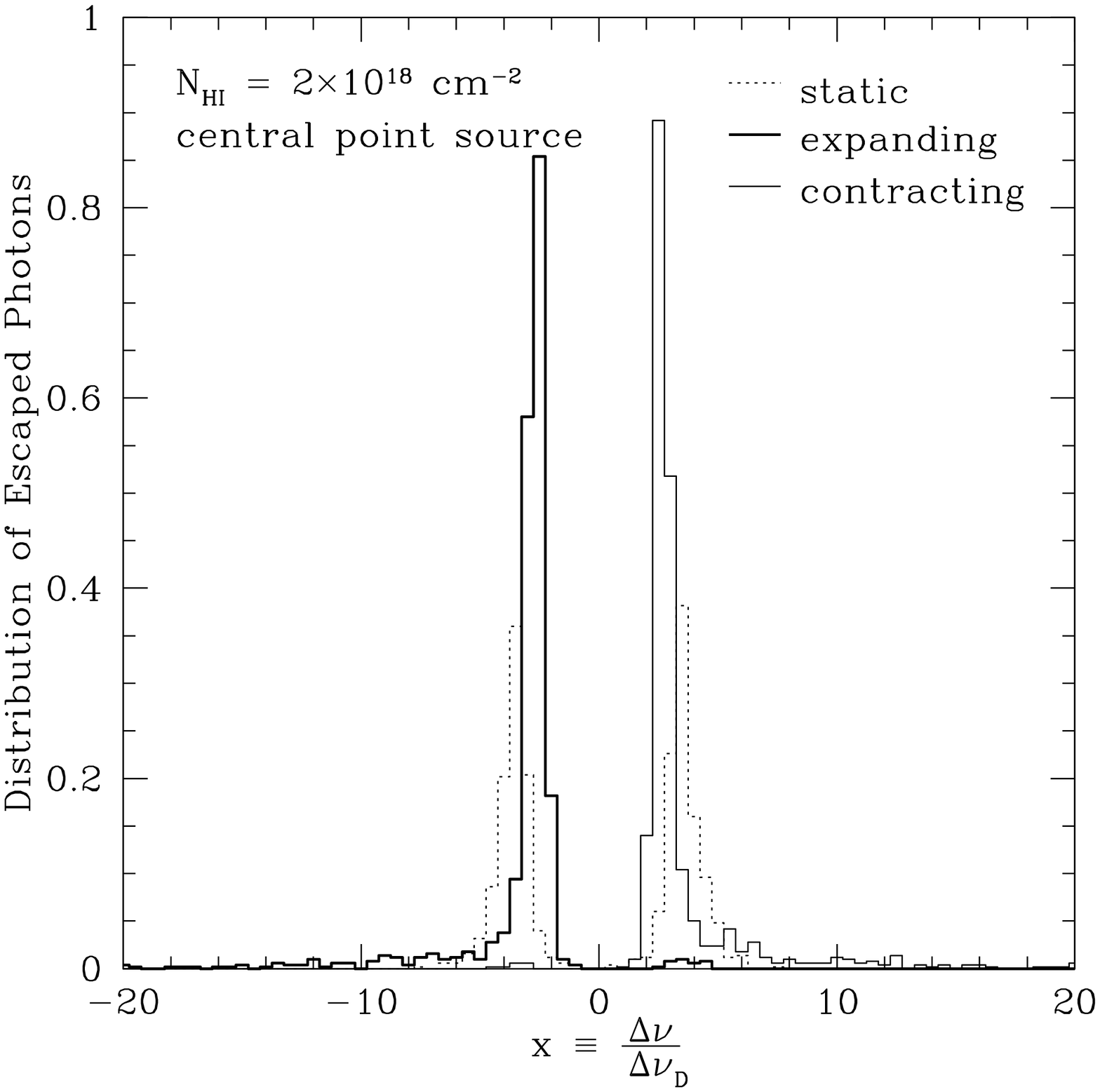,height=8cm,width=8cm} 
           }
\caption[figcases1]{\label{fig:cases1}
Frequency distribution of \lya photons escaped from a uniform hydrogen 
cloud with neutral hydrogen column density $2\times 10^{18} \cm^{-2}$
for the cases of uniform emissivity and central point source. Different
line types are for a static, expanding, or contracting cloud.
}
\end{figure}

\begin{figure}[h]
\centerline{\psfig{figure=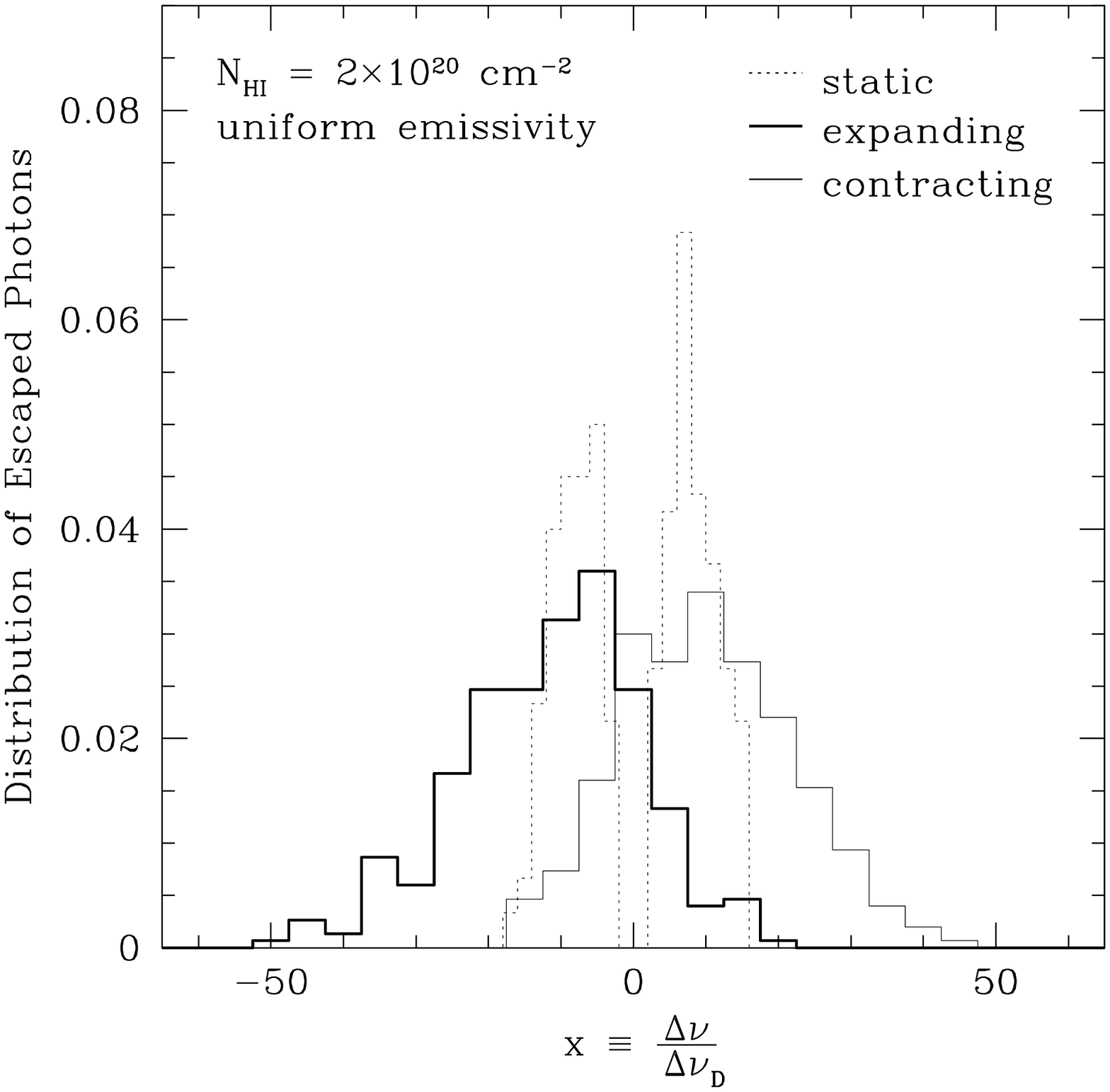,height=8cm,width=8cm}
            \psfig{figure=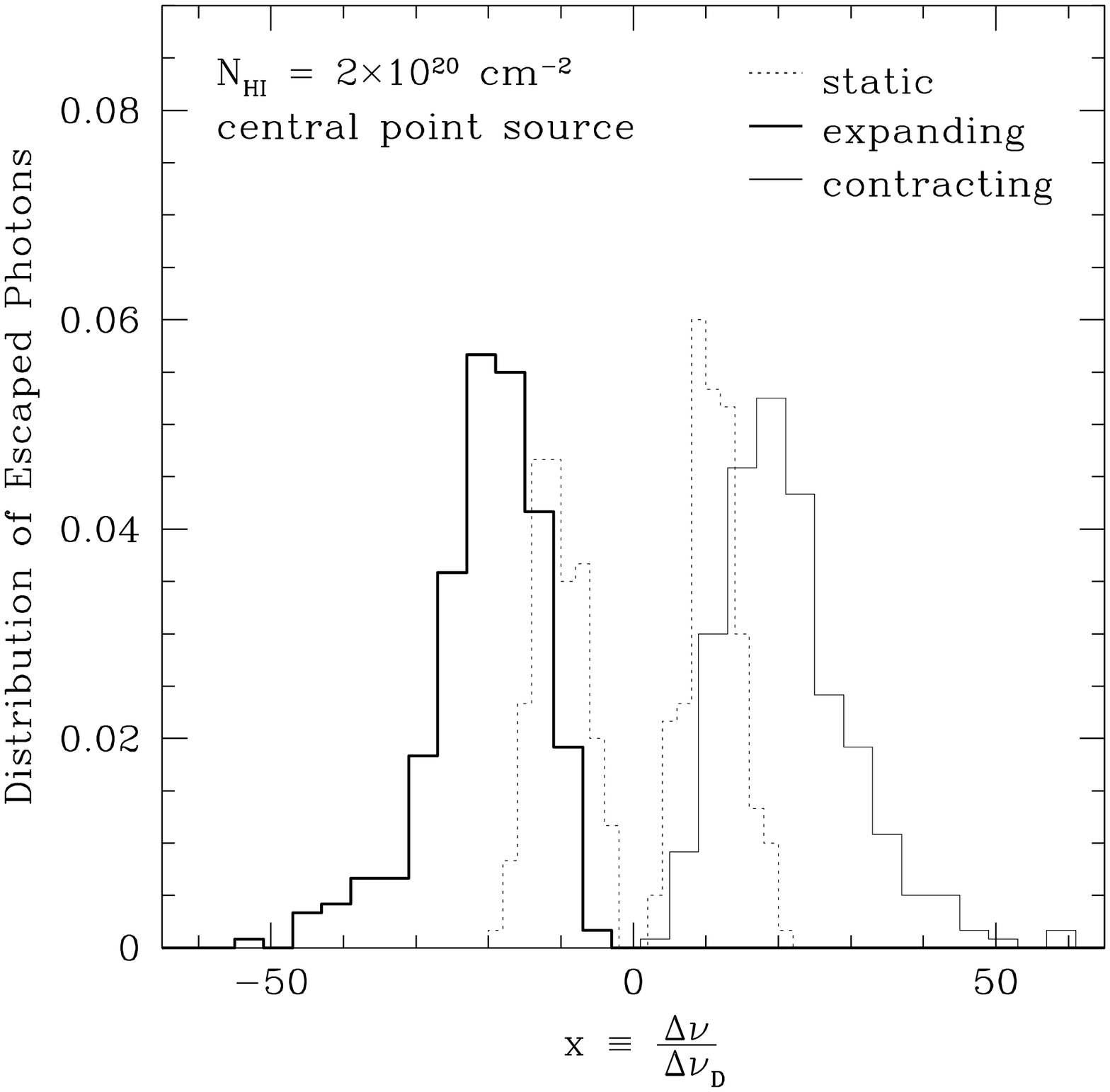,height=8cm,width=8cm} 
           }
\caption[figcases2]{\label{fig:cases2}
Same as Fig.~\ref{fig:cases1}, but for neutral hydrogen column density 
$2\times 10^{20} \cm^{-2}$.
}
\end{figure}

  We start applying our code to the most simple case of spherical
hydrogen clouds with uniform density. We consider the following cases:

Case 1: uniform emissivity, static cloud

Case 2: uniform emissivity, expanding or contracting cloud

Case 3: central point source, static cloud

Case 4: central point source, expanding or contracting cloud 

  For each case, we perform two runs with different column densities 
($2\times10^{18}\cm^{-2}$ and $2\times 10^{20}\cm^{-2}$, corresponding 
to line-center optical depths $\tau_0=8.3\times 10^4$ and $8.3\times 10^6$,
respectively). The temperature is assumed to be $2 \times 10^4$ K 
everywhere. In the case of expansion, we set it to be Hubble-like 
(i.e., the velocity is proportional to the radius), with the velocity 
at the edge of the system fixed to be $200 \kms$. The cases of 
contraction are done in the same way; their line spectra can generally 
be obtained by simply mirror-reflecting in frequency the photons escaped 
from the expanding cloud about the rest frame \lya frequency. 

  Figures \ref{fig:cases1} and \ref{fig:cases2} show the distribution of
escaped photons for all the cases. For static cases, as expected, the
escaped photons have a double-peaked frequency distribution. Except for 
the very small effect of atomic recoil (see Field 1959), the two peaks
are exactly symmetric; the differences in the figures are due to the
simulation noise.

  The results obtained for the emergent spectrum can be understood by 
noting the typical trajectory in frequency and space followed by a photon
(e.g., \citealt{Osterbrock62,Adams72,Urbaniak81,Ahn00}). In the case of 
a Lyman limit system (like in Fig.~2), the photons are always most likely 
to be scattered by atoms that have the right velocity along the photon 
direction so that, in their rest frame, the photon frequency is shifted 
very close to the line center. This implies there is practically no 
correlation between the frequency of the photon before and after scattering. 
At any random time, a photon is most likely to be found within a
frequency range $\Delta\nu_D=(v_p/c)\nu_0$ of the line center, and will
occasionally make larger excursions away from the line center when
scattered by an atom having a large velocity perpendicular to the
photon's direction. After any such excursion, the photon will likely
return immediately to the line center with one or a few scatterings.
Therefore, the photon escapes the cloud not by diffusing in frequency,
but by making a single large jump when scattered by an atom in the
high-velocity tail of the Maxwellian distribution, which reduces its
optical depth for escaping to a value of order unity. For a Lyman
limit system with $\nhi=2\times 10^{18}\cm^{-2}$, the line-center
optical depth is $\tau_0= 8.3\times 10^4$,
%f_{12} \pi^{1/2} e^2 \nhi /(m_e v_p \nu_0) = 8.3\times 10^4$, 
and the photons escape when their optical depth is
$\tau = \tau_0 e^{-x^2}\sim 1$ (where $x\equiv\Delta \nu/\Delta \nu_D$),
which implies $|x|\simeq \sqrt{\ln \tau_0} = 3.4$.

  In the case of a DLA system, the scattering history differs
from the Lyman limit case once the photon reaches a sufficiently large
excursion in frequency to make it more likely that the next scattering
is caused by a random atom far from the line center, rather than an atom
moving with the right velocity to shift the photon frequency to the line
center in the atom frame. Then, the evolution of the photon is
described by diffusion in frequency. Since the photons are now reaching 
their escaping frequency not by a large jump but by a series of
small steps, they undergo greater spatial diffusion than in Lyman limit
systems. This, plus the fact that the optical depth has now a power-law
instead of Gaussian dependence at large $x$, broadens the width of the
two emission peaks. Of course, the emission peaks also move further from
the line center because a smaller scattering cross section is required
for the photons to escape.

\begin{figure}[h]
\centerline{\psfig{figure=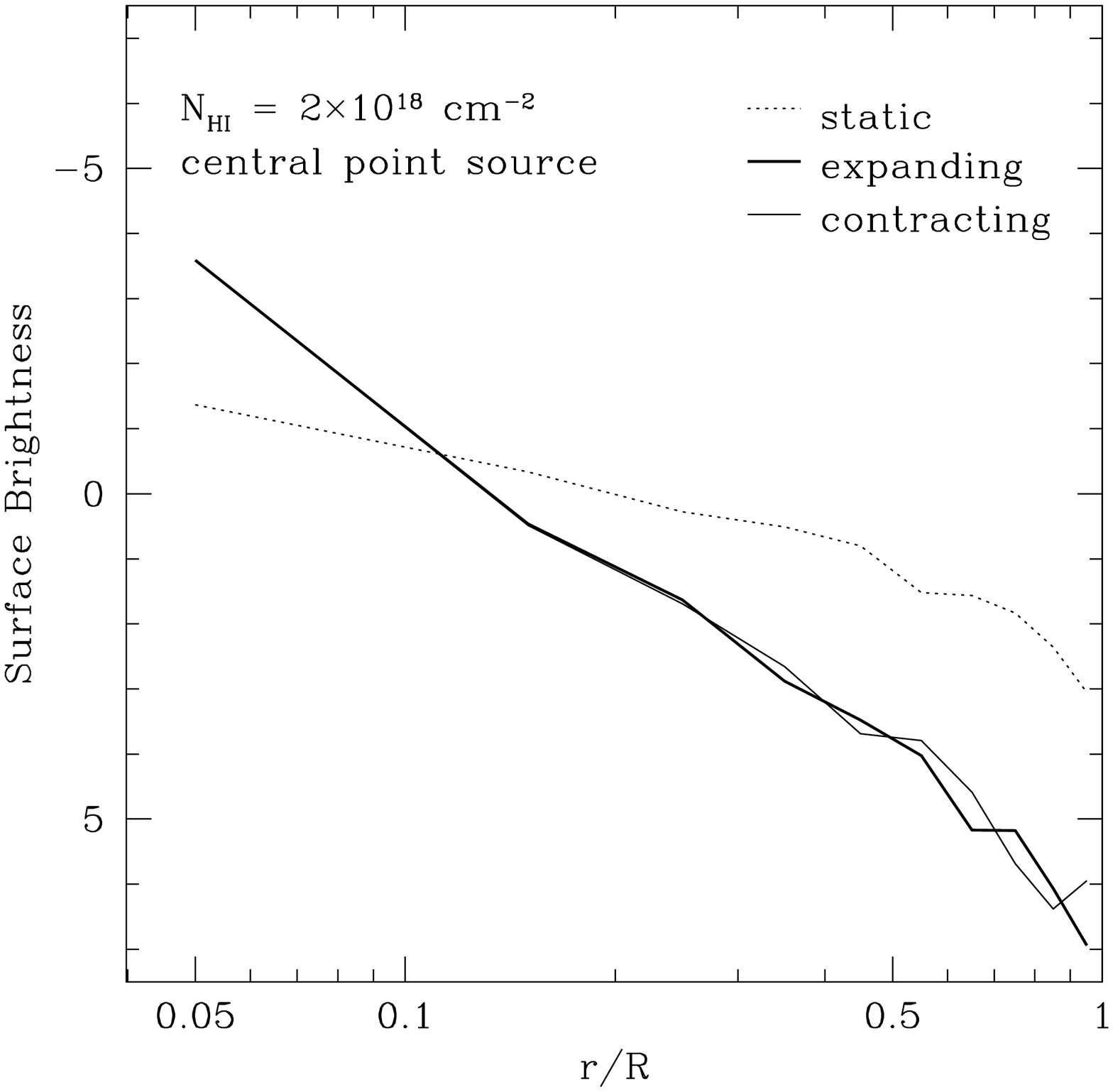,height=8cm,width=8cm}
            \psfig{figure=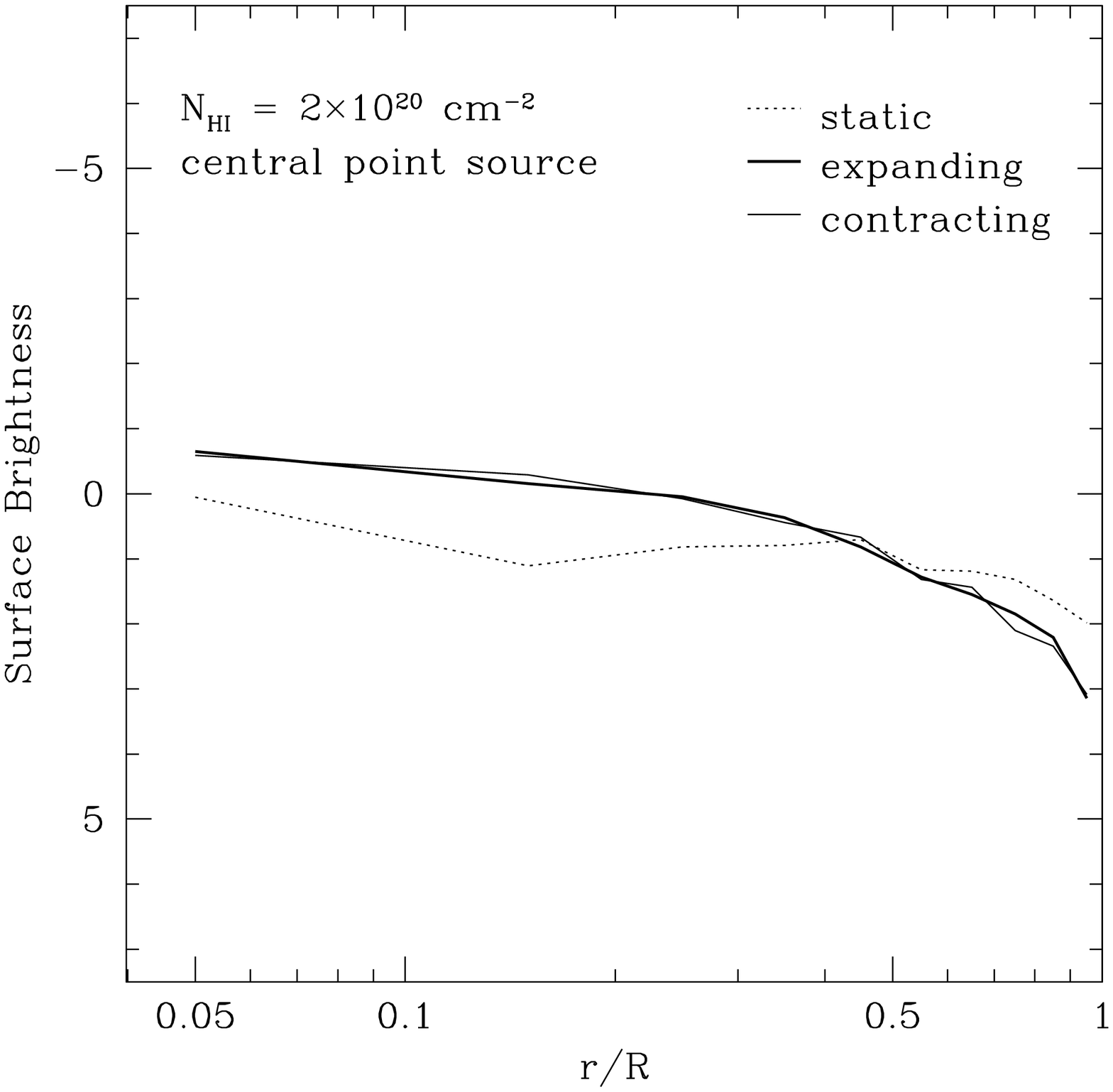,height=8cm,width=8cm} 
           }
\caption[figsbp]{\label{fig:SBP0}
Surface brightness profiles, in magnitudes per solid angle (with an
arbitrary zero point). Curves are normalized in such a way that the
total energy of the photons remains the same. In the right panel, the 
drop at $r/R \sim 0.15 $ for the static case is due to simulation noise.
}
\end{figure}

  In Figure \ref{fig:SBP0}, we can see that the surface brightness produced
by a central point source is more extended for a DLA system than
for a Lyman limit system, owing to the greater spatial diffusion.

  For an expanding cloud, photons should escape on average with a redshift
because they are doing work on the expansion of the cloud as they are
scattered. For a Lyman limit system with a central source, a photon
undergoing a large positive frequency jump will be moved back to the
line center relative to the fluid as it travels through the cloud, owing
to the Hubble-like expansion of the cloud. On the other hand, a negative 
frequency jump will allow the photon to escape directly. In our case, the 
velocity at the cloud edge, $200 \kms$, is much larger than the atomic 
velocity dispersion, so a photon needs to undergo many positive frequency 
jumps in order to diffuse spatially through the cloud and be able to escape 
on the blue side of the line, which explains why the blue peak is highly
suppressed (see Fig.~3 of \citealt{Ahn98} for a similar explanation). 
The situation is exactly reversed for a contracting cloud.
The case of uniform emissivity broadens the line, essentially because
of the different velocities of the emission sites of the photons; there
is also the additional effect that photons emitted near the edge of the
cloud are more likely to escape on the blue (red) peak for an expanding
(contracting) cloud. In DLAs, the line is also broader
for a point source compared to a Lyman limit system because of the
power-law dependence of the cross section on frequency and the greater
degree of spatial diffusion.

  To summarize, these simple models show how the frequency and spatial 
distributions of \lya photons escaped from a cloud are related to the
\lya emissivity distribution, the bulk velocity field and the column
density of the cloud.
 
\section{\lya Emission from Model Damped \lya Systems}

  We now apply our code to a gas cloud with an isothermal density
profile as a model for DLAs. The nature of DLAs is still a subject of
debate. They could be protogalaxies with a rotating disk component
(e.g., \citealt{Prochaska97,Prochaska98}) or spherically distributed clouds
of gas moving randomly in halos (e.g., \citealt{Mo94,Haehnelt98,McDonald99}). 
The other mechanism that may give rise to DLAs is large-scale outflows due 
to galactic winds in dwarf galaxies (e.g., \citealt{Nulsen98,Tenorio99,
Schaye01}). In this paper, we focus on the first two pictures. 
We investigate the emergent spectra and spatial distribution of the 
\lya emission, to show what it can tell us about the internal structure 
of the system.

\subsection{Model Description}

  We model the gas cloud producing a DLA assuming that it forms in a dark
matter halo of mass $10^{11}M_\odot$, virialized at redshift $z=3$. In
an $\Omega_M=1$ universe with $H_0=70 \, {\rm km\,s^{-1}Mpc^{-1}}$, the
corresponding virial radius is $r_{\rm vir}=24 \kpc$, and the virial
velocity $V_{\rm vir}=104 \kms$ (e.g., \citealt{Padmanabhan93}). We assume 
the gas density profile is a singular isothermal sphere with a cutoff at
the virial radius and a fraction of the halo mass in gas of 0.05.  

  We consider two different cases as an illustration to show how the
rotation of a system could be probed by observations of \lya emission.
The first case is a spherically symmetric cloud, and the second a
rotating oblate ellipsoid with an axis ratio of $0.5$ for the gas
distribution. The rotating velocity is set to be $V_{\rm rot} =
\sqrt{2/3}V_{\rm vir}$. In one case, we assume the gas to be static
and at a temperature $2\times10^4$ K (a typical temperature of gas that
cools after shock-heating but stays photoionized). We also consider
another case where the gas is given an additional velocity dispersion
of $V_{\rm vir}/\sqrt{3}$ in the spherical model, and $V_{\rm vir}/3$ in the
flattened, rotating model. This additional velocity dispersion is
simply added quadratically to the thermal one, which would be valid if
the gas were in optically thin clumps moving at this velocity
dispersion. In practice, any relevant gas clumps in DLAs will be
optically thick to \lya photons, but their inclusion would make our
model much more complex. We will discuss the effect we would expect from
optically thick clumps in \S 5.

  There are various sources to produce \lya photons: internal dissipation,
star formation, and fluorescence caused by the intergalactic UV background.
The external UV background will give rise to \lya fluorescence; at 
$\sim 10^4$ K, in an optically thick medium, recombination of the 
photoionized hydrogen has a 68\% probability of producing a \lya photon 
\citep{Spitzer78}. Shock-heating of the gas
will dissipate kinetic energy into heat, which will result in \lya
emission after line excitation and collisional ionization followed by
recombination. In addition, internal star formation can of course also
produce \lya emission. We consider only fluorescent emission from the
external background in this paper. In this case, the emissivity 
is proportional to the square of the ionized gas density. The other two 
sources of emission are more centrally concentrated, and therefore similar 
differences as the ones between uniform emissivity and central source in 
the previous section can be expected.

  We take self-shielding into account in the spherical case,
calculating the neutral hydrogen density of the cloud at every radius,
where the total gas density profile is singular isothermal. We use an
iterative algorithm similar to the one presented by Tajiri \& Umemura
(1998), as described in \citet{Zheng02}. Here, we
assume a UV background intensity of 
$10^{-22} \ergs\cm^{-2}\seg^{-1}\hz^{-1}\ster^{-1}$,
constant at all frequencies between the \ion{H}{1} Lyman limit, $\nu_L$,
and the \ion{He}{2} Lyman limit, $4\nu_L$. The intensity is set to zero
at frequencies above $4\nu_L$. This is a good approximation since 
most higher frequency photons are likely to be absorbed by \ion{He}{2} before 
reaching the hydrogen self-shielded zone. Even though in reality some 
very energetic photons exist, which can penetrate into the self-shielded 
region, their effect on the profile of neutral hydrogen is very small 
due to their low intensity.  The profile of neutral hydrogen that is
obtained is similar to that shown in \citet{Zheng02}. In the case of
the flattened, rotating gas distribution, we simply take the same
neutral hydrogen density profile as for the spherical case and flatten
it with an axis ratio of $0.5$ (the numerical code we have developed for
computing the self-shielding correction applies only to spherically
symmetric systems). This is of course not exact, although a much better
approximation than neglecting self-shielding altogether.
 
\subsection{Results of Simulations}

\begin{figure}[h]
\centerline{\psfig{figure=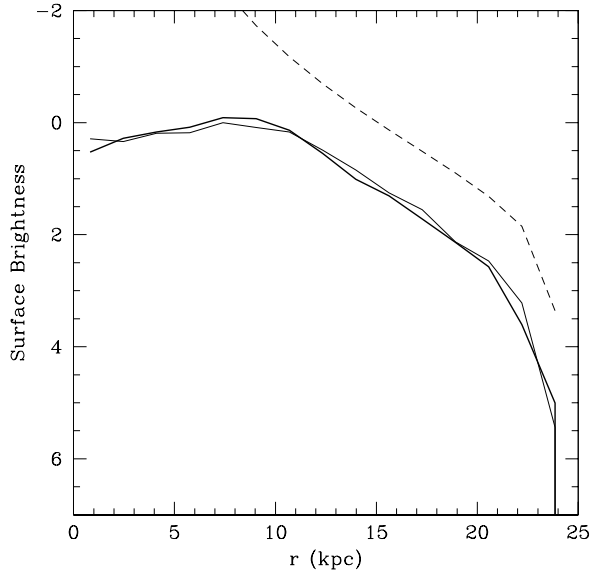,height=8cm,width=8cm}}
\caption[]{\label{fig:sbp}
Surface brightness profiles of \lya emission for the spherically
symmetric cloud (in magnitudes per unit solid angle, with an arbitrary
normalization). Thick and thin lines are with and without velocity
dispersion, respectively (see the text). Dashed line is for optically 
thin cloud (with an arbitrary vertical shift).
}
\end{figure}

  The surface brightness profile for the spherically symmetric,
non-rotating cloud is plotted in Figure \ref{fig:sbp}. The thick
solid line shows the case of including the fluid velocity dispersion,
$\sigma=V_{\rm vir}/\sqrt{3}=60 \kms$, while the thin solid line is the
result with only the thermal velocity dispersion, $(kT/m_H)^{1/2} =
12.8 \kms$. The dashed line is what is obtained when self-shielding
is not included (it has been shifted vertically). The velocity dispersion 
has practically no effect on
the surface brightness profile, because the change in spatial diffusion
of the photons is very small. The emissivity due to the recombination
has a ``hole" in the inner part of the cloud because of self-shielding.
Therefore, we see a ``core" in the surface brightness profile (with a
slight decline of surface brightness toward the center in the inner
part due to geometric effects). The outer part of the profile follows
the optically thin case. The core of the surface brightness profile is a
signature that the emission is due to fluorescence of the external
background, since other sources of emission are more centrally
concentrated.

  Because the ionizing photons of the external background extend only up
to $4 \nu_L$ in frequency, essentially all the photoionizations and 
subsequent recombinations occur in the outer region of the gas cloud with 
a hydrogen column density of $\nhi \lesssim 10^{19}\cm^{-2}$, the inverse 
of the photoionization cross section at frequency $4\nu_L$. The damped 
absorption wings are not yet important at this column density, therefore 
the photons escape the cloud after a single large jump in frequency when 
scattered by an atom in the high-velocity tail of the Maxwellian 
distribution, as discussed in \S 3 for the case of Lyman limit systems.  

\begin{figure}[h]
\centerline{\psfig{figure=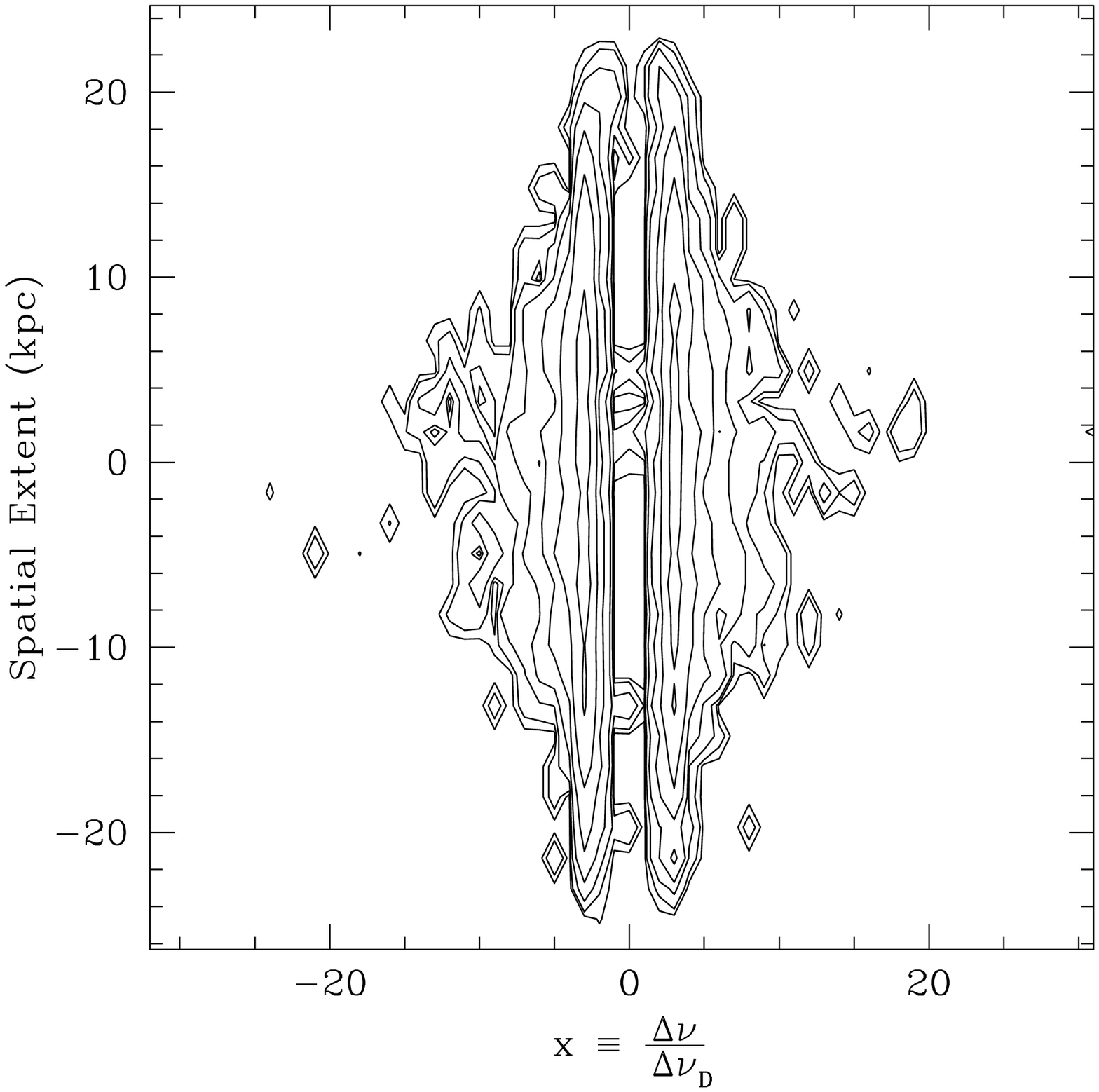,height=8cm,width=8cm}
            \psfig{figure=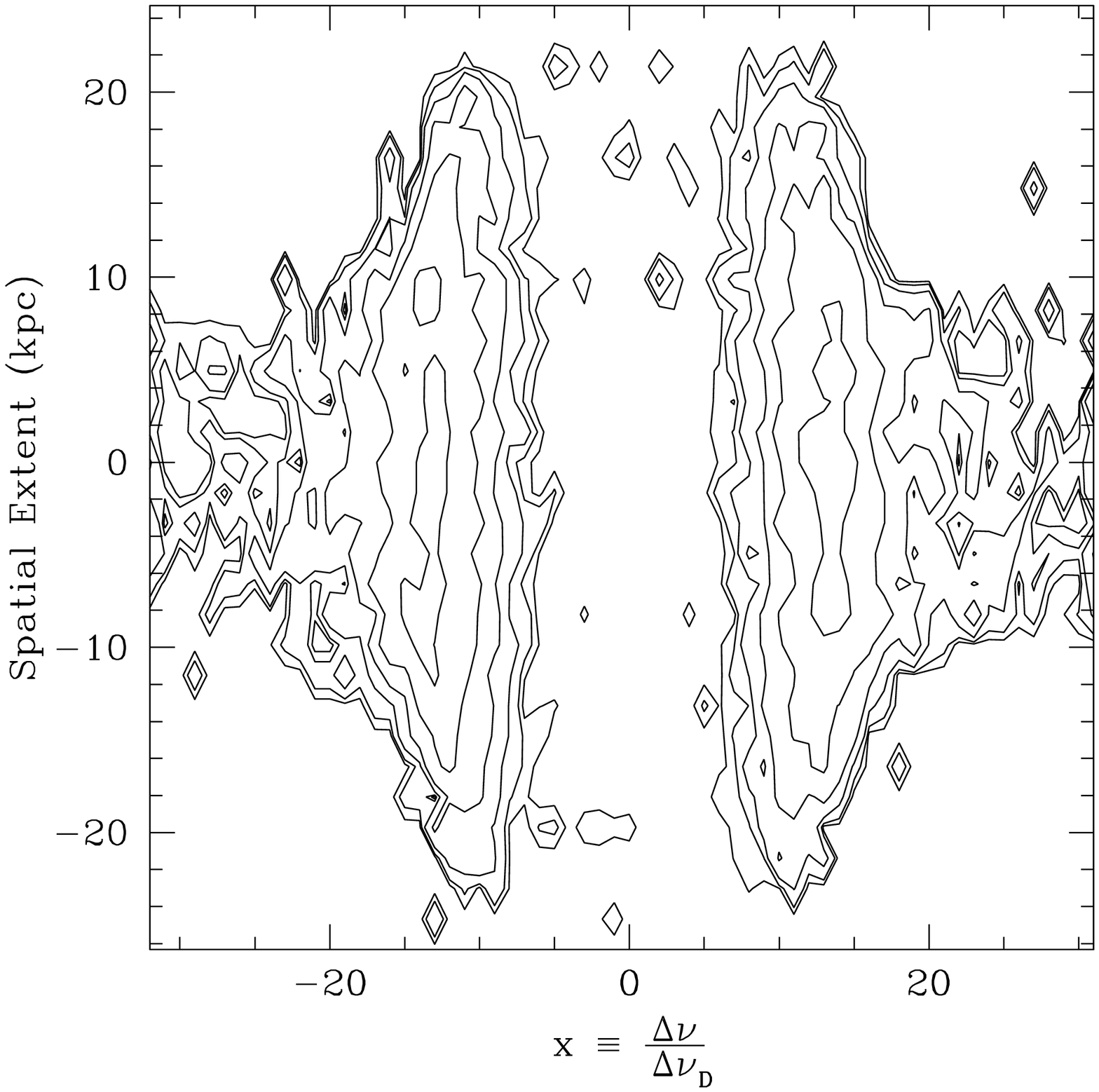,height=8cm,width=8cm}
           }
\caption[]{\label{fig:spec2dstat}
Two-dimensional spectra of the spherically symmetric, non-rotating cloud.
Left panel is for a thermal velocity dispersion of $(kT/m_H)^{1/2} =
12.8 \kms$ only, right panel includes a fluid velocity dispersion of
$60 \kms$ (see the text). The contours next to each other are separated 
by 1 magnitude.
}
\end{figure}

  Figure \ref{fig:spec2dstat} shows the 2-dimensional spectra of the 
\lya photons escaped from the non-rotating cloud. We assume that the 
slit is large enough to include the entire cloud. The double-peaked
distribution of escaped \lya photons, with the intermediate frequency
range of essentially zero flux, appears at every spatial position.

  Including an additional velocity dispersion has the effect of
separating the two peaks. This is true in our model because we consider
the case where there is no correlation between the velocity of two atoms
that are spatially close, valid only when any gas clumps moving
coherently are so small that they are optically thin to \lya photons.
In other words, the additional velocity dispersion is effectively
thermal. In the presence of coherent motions of optically thick regions,
the intermediate frequency range with a highly reduced flux disappears,
as discussed in \S 3 in the examples of an expanding or contracting
cloud.

\begin{figure}[h]
\centerline{\psfig{figure=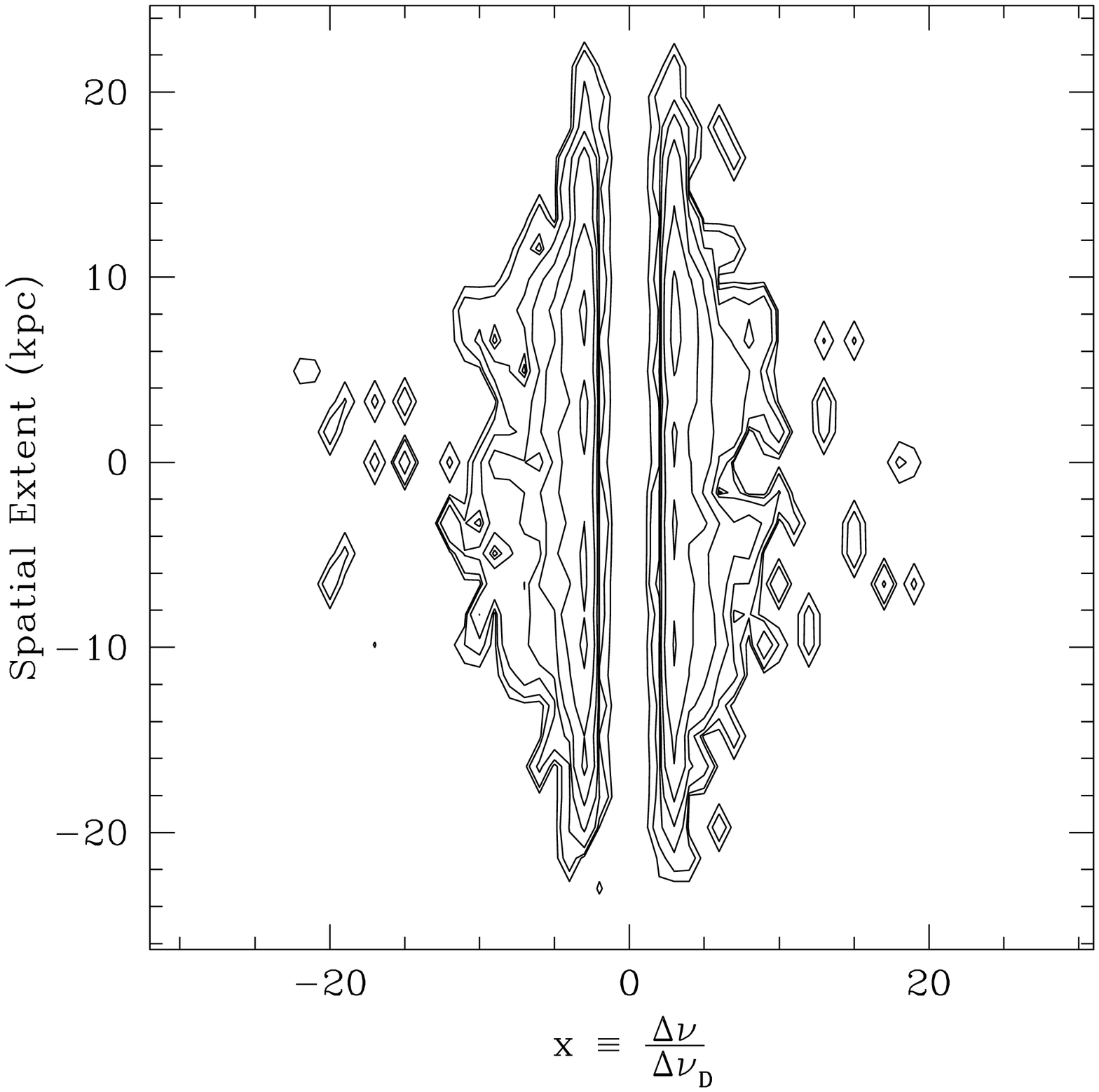,height=8cm,width=8cm}
            \psfig{figure=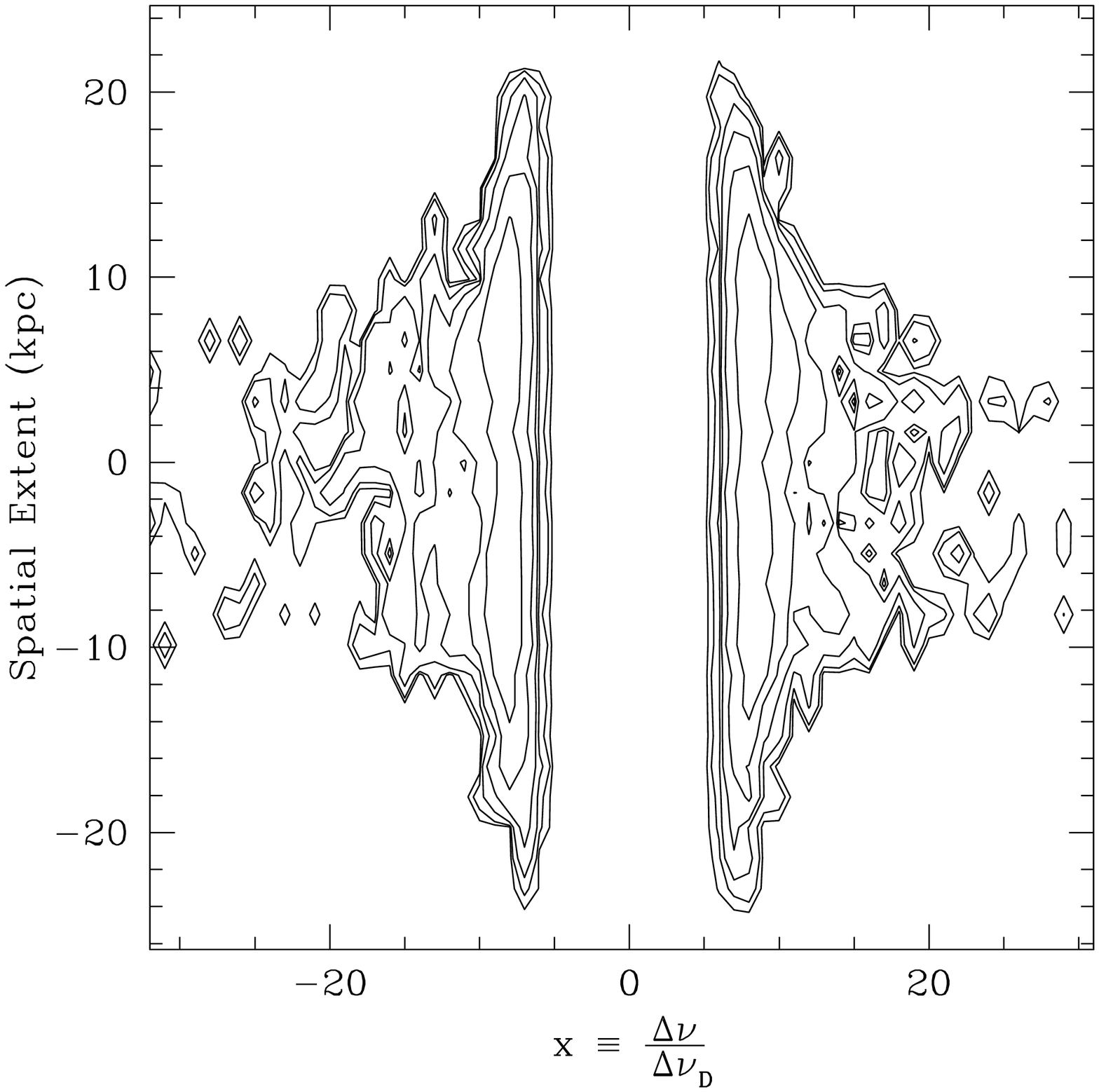,height=8cm,width=8cm}
           }
\caption[]{\label{fig:spec2drot1}
Two-dimensional spectra of the oblate rotating cloud viewed face-on.
Left panel is for a thermal velocity dispersion of $12.8 \kms$ only, 
right panel includes an additional velocity dispersion of $35 \kms$ 
(see the text). The contours next to each other are separated 
by 1 magnitude.
}
\end{figure}

\begin{figure}
\centerline{\psfig{figure=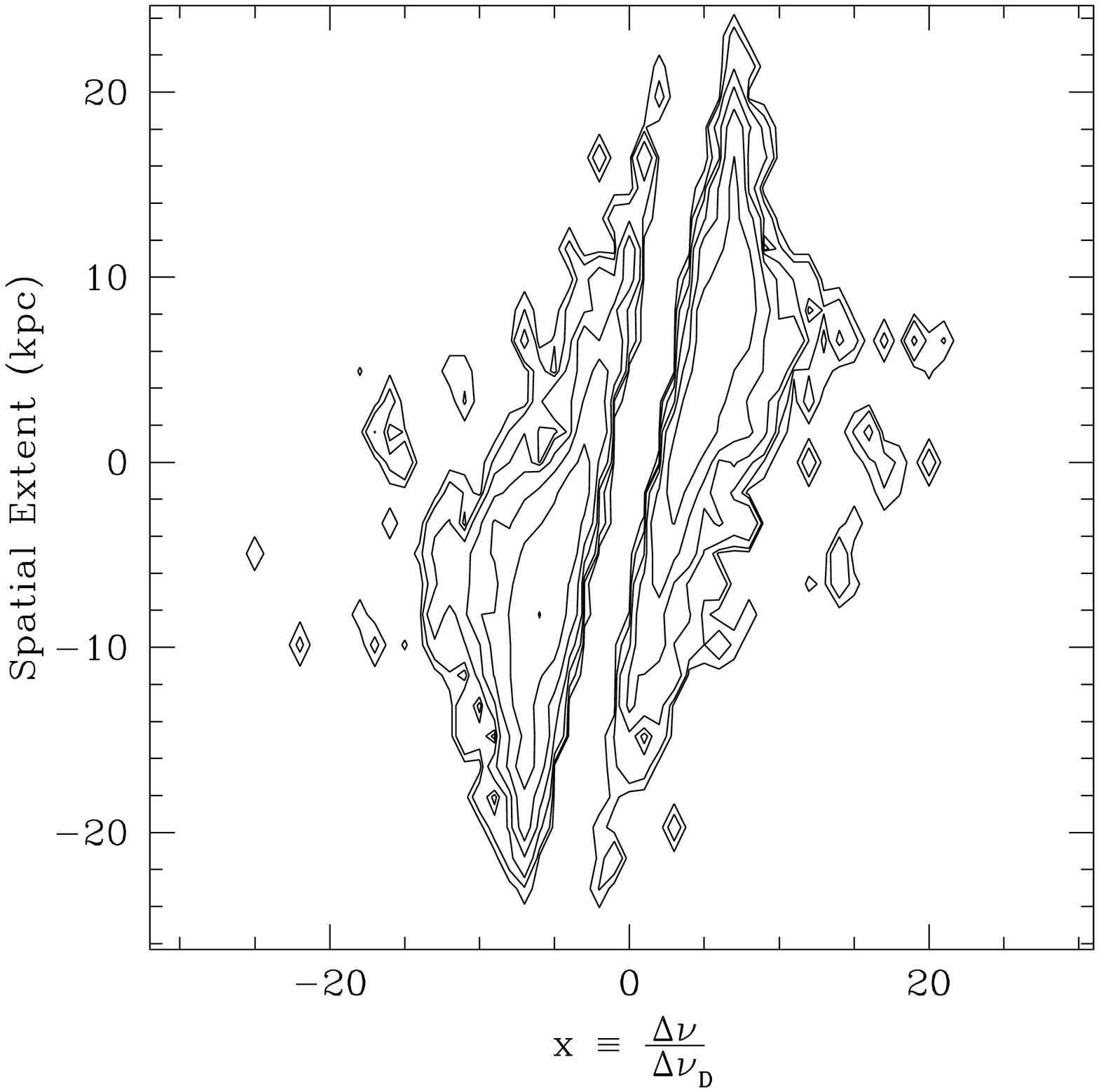,height=8cm,width=8cm}
            \psfig{figure=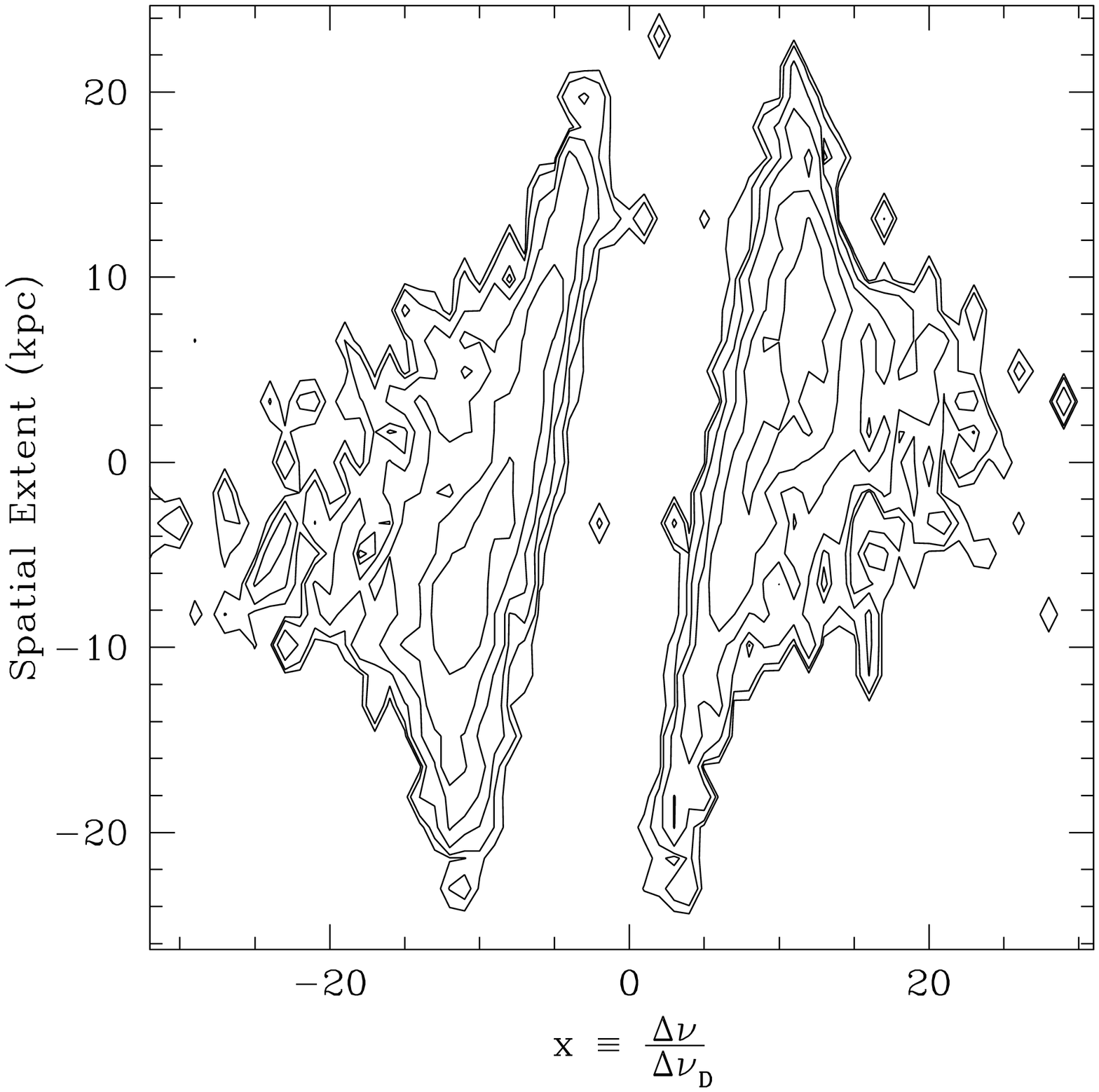,height=8cm,width=8cm}
           }
\centerline{\psfig{figure=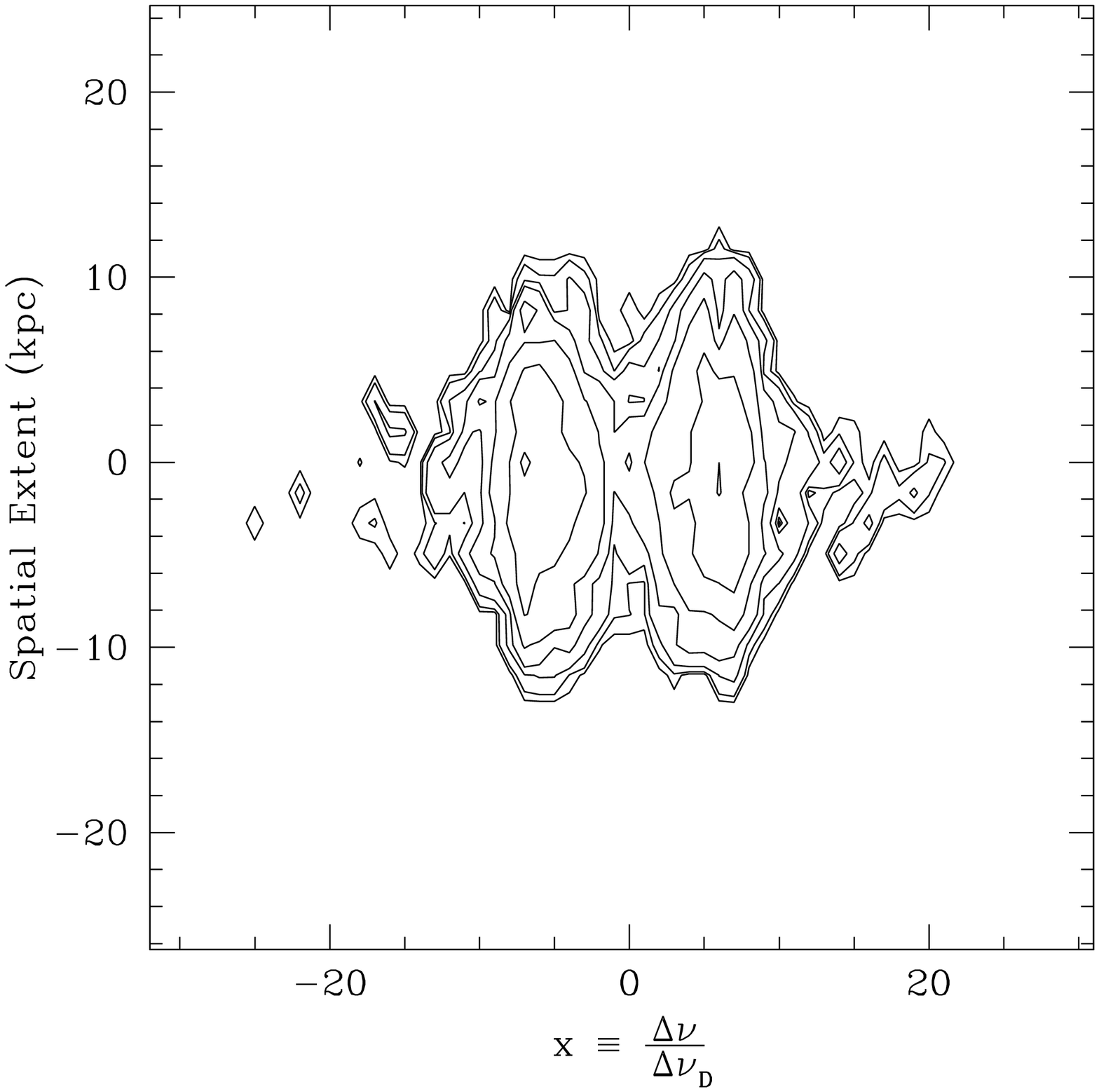,height=8cm,width=8cm}
            \psfig{figure=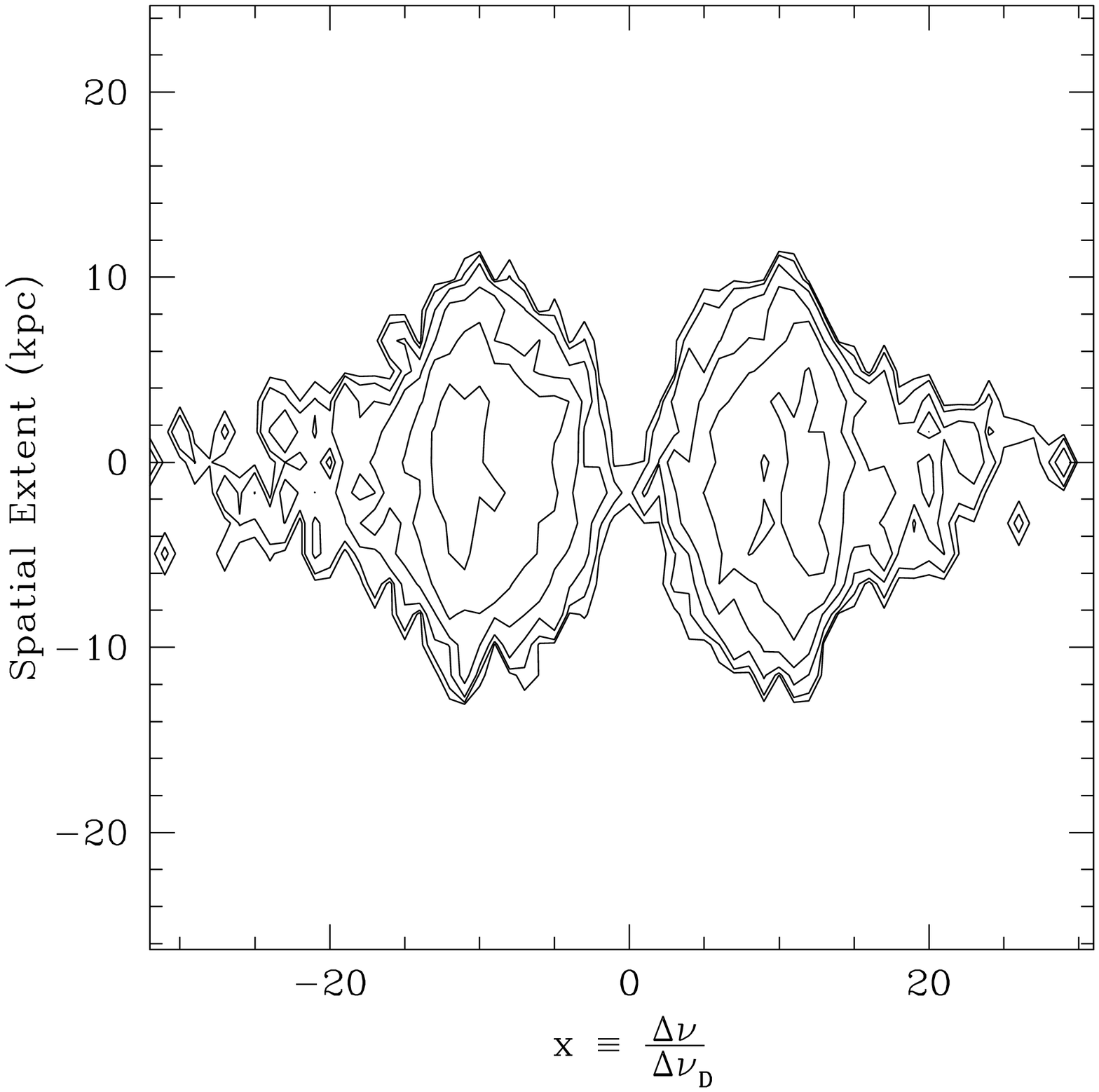,height=8cm,width=8cm}
           }
\caption[]{\label{fig:spec2drot2}
Two-dimensional spectra of the oblate rotating cloud viewed edge-on, with
the slit perpendicular (top panels) and parallel (bottom panels) to the
rotation axis. Left and right panels are for cases of different velocity
dispersions as explained in Fig.~7. The contours next to each other 
are separated by 1 magnitude.
}
\end{figure}

  Spectra of the flattened rotating cloud depend on the viewing angle
and slit orientation. The face-on spectrum, plotted in Figure
\ref{fig:spec2drot1}, is very similar to that of the spherical static case.
Figure \ref{fig:spec2drot2} shows the edge-on spectra, with the slit
perpendicular (top panels) and parallel (bottom panels) to the
rotational axis. The spectrum is averaged along the spatial coordinate
perpendicular to the slit; in other words, we are assuming that the
entire cloud is included in the slit, but that the wavelength dispersion
is wide enough that the size of the cloud does not introduce any
smoothing of the spectrum. When the slit is perpendicular to the
rotation axis, the rotation curve pattern is clearly seen. The velocity
shift of the spectrum is about the same as the rotational velocity,
$V_{\rm rot}=V_{\rm vir}\sqrt{2/3}= 85 \kms$, at the edge of the system
[note that the unit used in the horizontal axis is $\Delta \nu_D = 
(2kT/m_H)^{1/2} \nu_0/c = 18.2 \kms (\nu_0/c)$]. The variation of the
shape of the spectrum along the slit is easily understood by thinking
of an analogy to the expanding and contracting cloud in \S 3. The
\lya photons are produced on the outer ionized layer of the system,
near the radius $r_{\rm ss}$ where the hydrogen becomes self-shielding. 
The line-of-sight velocity at this radius is proportional to 
$V_{\rm rot} r_p/r_{\rm ss}$, where $r_p$ is the projected radius on the 
slit position. Therefore, the central trough of the spectrum shifts 
linearly with the projected radius. Moreover, the variation of the 
velocity along the line of sight causes the peak of emission that is 
further from the mean velocity of the system to be enhanced, for similar 
reasons as in the case of the expanding and contracting clouds discussed 
in \S 3. 

  For the case of the slit parallel to the rotational axis, the
2-dimensional spectrum displays a symmetric double-peaked pattern,
which essentially results from averaging along the equator the spectrum
with different projected velocity. This averaging results in a less
sharp central trough and smoothing of the two peaks. In all cases,
the velocity dispersion increases the distance between the two peaks in
the photon distribution and broadens the width of each peak.  

\section{Summary and Discussion}

  We have developed a Monte Carlo code of \lya photon scattering, to
construct simulated images and spectra. The code is fully
three-dimensional and adaptable to a cloud with any given \lya
emissivity, neutral hydrogen density distribution, and bulk velocity
field. Several simple cases have been presented to show the images and
spectra expected from a uniform gas distribution. We have applied the
code to model DLA systems illuminated by the intergalactic UV background
to study the spatial and frequency distribution of emitted \lya photons.
Self-shielding produces a core in the \lya surface brightness profile.
For a spherical, static cloud, \lya photons have a double-peaked
distribution which is symmetric around the line center frequency. An
oblate, rotating cloud shows the same double-peaked distribution with
the rotational pattern of increasing mean wavelength with the position
on the slit. Observations of the fluorescent emission from damped \lya
systems with large telescopes (\citealt{Hogan87,Gould96}) could reveal 
the presence of such systems and measure a velocity
gradient, providing a direct measurement of the rotation rate. The \lya
image would also reveal if the emission is due to fluorescence from
the external radiation (in which case we expect a large core of the
surface brightness at the radius where the hydrogen becomes
self-shielded) or due to internal gas dissipation or star formation.
  
  The observations of metal lines in DLA systems show multiple
absorption lines (e.g., \citealt{Prochaska97,Prochaska98}), suggesting 
that the gas is clumpy. In this paper, we have assumed instead that the 
gas has a smooth distribution. The case of enhanced velocity dispersion 
we have considered is valid only if the gas is in very small clumps that
are optically thin to \lya photons. Clumps that can give rise to the
observed metal lines are highly optically thick. The effect of a clumpy
gas distribution is not difficult to imagine. If there is only one
clump along a line of sight, then this clump will essentially produce
the same spectral shape in its emission line, shifted to its
velocity. Thus, if the emission line could be observed with sufficient
angular resolution to resolve the individual clumps, the characteristic
spectral shape of Figure \ref{fig:spec2dstat} would appear at every 
position, although shifted to the clump velocity. If the clumps are not 
resolved, the emission line shape is obviously smoothed, just like in the 
case of an expanding or contracting cloud shown in Figure \ref{fig:cases1}. 
Our numerical code can be applied in the future to the results of detailed 
hydrodynamic simulations of gaseous halos in the process of forming 
galaxies in cosmological realizations to predict other observable 
signatures that can test the mechanisms by which energy dissipation of gas 
leads to galaxy formation.

A possible effect that we have not considered in this paper is the 
absorption of \lya photons by dust. Dust seems to be present in at least 
some DLAs as indicated by the reddening of the background quasars that 
have DLAs along the line of sight (\citealt{Fall89,Pei91}) and the abundance 
ratios of iron-peak elements which follow a dust depletion pattern 
(\citealt{Savage96,Vladilo98}). The resonant nature of the scattering of 
\lya photons by neutral hydrogen atoms increases the probability 
for them to be absorbed by dust grains. This is probably relatively more 
important when the \lya emission originates from a star-forming region 
(e.g., \citealt{Meier81,Charlot93}), where dust is likely to be abundant. 
If \lya photons of DLAs are caused by fluorescent emission as cases we 
consider in this paper, the absorption by dust would not be very severe due 
to the low column densities of the ionized outer layer \citep{Charlot91}. 
Since in the presence of dust the escape of \lya photons depends on many 
factors, such as the \ion{H}{1} column density, the distribution of \lya 
source, the property of dust, and the topology of the medium (e.g., 
\citealt{Ferland79,Hummer80,Neufeld90, Neufeld91,Charlot91}), detailed 
modeling of DLAs is necessary for dust absorption to be taken into 
consideration.

\acknowledgments

The authors thank the referee Joop Schaye for helpful comments. 
We also thank Xuelei Chen and David Weinberg for useful discussions.
This work was supported in part by NSF grant NSF-0098515.

%\clearpage
\bigskip
\bigskip

\centerline{\appendix{\bf APPENDIX}}

To choose the velocity along the direction of the incident photon
of the atom responsible for the scattering, we need to generate a
random number $u$ that has the following distribution (see eq.[4]):
$$ f(u) \propto \frac{e^{-u^2}}{(x-u)^2+a^2}, \eqno({\rm A1})$$
where $x$ and $a$ are given. We give a brief description on how we 
generate random numbers with this distribution. Since the distribution 
has the symmetry under the transformation of $(x,u) \rightarrow (-x,-u)$, 
we limit the following discussion to $x>0$.

We make use of the rejection method \citep{Press92} to generate $u$. The 
comparison distribution can be chosen to be 
$g(u) \propto [(x-u)^2+a^2]^{-1}$, which can be integrated and inverted 
analytically. A value of $u$ with the 
distribution $g(u)$ is first generated. We keep this value only if a second 
random number uniformly distributed between 0 and 1 is smaller than 
$e^{-u^2}$. In this way, we obtain values of $u$ with the distribution
$f(u)$. In practice, when $x \gg 1$, the above method is inefficient because 
a very small fraction of values of $u$ are not discarded. To increase the
fraction of acceptance, we modify the comparison distribution to be
$$
g(u) \propto \left\{\begin{array}{ll}
             [(x-u)^2+a^2]^{-1},           &  u \leq u_0 \\
             e^{-u_0^2}[(x-u)^2+a^2]^{-1}, &  u > u_0 
             \end{array}
      \right.
\eqno({\rm A2})
$$
The value of $u_0$ can be chosen to minimize the fraction of
generated values that will be discarded. The acceptance fractions are
then required to be $e^{-u^2}$ and $e^{-u^2}/e^{-u_0^2}$ in the
regions $u \leq u_0$ and $u > u_0$, respectively. 

A first random number $R$ uniformly distributed between 0 and 1 
determines which region we use by comparing it with $p$, where
$$ p=\frac{\int_{-\infty}^{u_0} g(u) du}{\int_{-\infty}^{+\infty} g(u) du}
    =(\theta_0+\frac{\pi}{2})[(1-e^{-u_0^2})\theta_0+(1+e^{-u_0^2})
     \frac{\pi}{2}]^{-1},~ \theta_0=\tan^{-1}\frac{u_0-x}{a}. 
\eqno({\rm A3})
$$ 
We generate $u$ through $u=a\tan\theta+x$, where $\theta$ is a random 
number uniformly distributed in $[-\frac{\pi}{2},\theta_0]$ and 
$[\theta_0,\frac{\pi}{2}]$ for $R \leq p$ and $R > p$, respectively. 
Then another random number uniformly distributed between 0 and 1 determines
whether the generated value of $u$ is accepted by comparing it with the
corresponding fraction of acceptance.

\end{document}